\documentclass[%
reprint,
superscriptaddress,
amsmath,amssymb,
aps,
prb,
]{revtex4-2}
\usepackage{lipsum}
\usepackage[dvipsnames]{xcolor}
\usepackage{graphicx}% Include figure files
\usepackage[caption=false]{subfig}
\usepackage{dcolumn}% Align table columns on decimal point
\usepackage{bm}% bold math
\usepackage{hyperref}% add hypertext capabilities
\hypersetup{
  colorlinks,
  citecolor=blue,
  linkcolor=blue,
  urlcolor=blue}
\usepackage{color}
\usepackage{booktabs}

\newcommand{\beginsupplement}{
        \setcounter{table}{0}
        \renewcommand{\thetable}{S\arabic{table}}
        \setcounter{figure}{0}
        \renewcommand{\thefigure}{S\arabic{figure}}
        \setcounter{equation}{0}
        \renewcommand{\theequation}{S\arabic{equation}}
        \setcounter{page}{1} % Reset page counter to 1
        \renewcommand{\thepage}{S\arabic{page}} % Change page numbering style for Supplementary Information
        \onecolumngrid
}

\begin{document}

\preprint{APS/123-QED}
\title{Digital Quantum Simulation of Spin Transport}

\author{Yi-Ting Lee}
\affiliation{Department of Materials Science and Engineering, University of Illinois at Urbana-Champaign, Urbana, IL 61801, USA}

\author{Bibek Pokharel}
\affiliation{IBM Quantum, IBM Thomas J. Watson Research Center, Yorktown Heights, NY, USA}
\affiliation{Quantum Science Center, Oak Ridge National Laboratory, Oak Ridge, TN 37831, USA}

\author{Jeffrey Cohn}
\affiliation{Quantum Science Center, Oak Ridge National Laboratory, Oak Ridge, TN 37831, USA}
\affiliation{IBM Quantum, IBM Research – Almaden, San Jose, CA 95120, USA}

\author{Andr\'e Schleife}
% \email{schleife@illinois.edu}
\affiliation{Department of Materials Science and Engineering, University of Illinois at Urbana-Champaign, Urbana, IL 61801, USA}
\affiliation{Materials Research Laboratory, University of Illinois at Urbana-Champaign, Urbana, IL 61801, USA}

\author{Arnab Banerjee*}
\affiliation{Quantum Science Center, Oak Ridge National Laboratory, Oak Ridge, TN 37831, USA}
\affiliation{Department of Physics and Astronomy, Purdue University, West Lafayette, IN 47906, USA}

% well, i don't know what is the exact order of this, put i put * on arnab as he is the corresponding author

% \date{\today}
% \yt{Feel free to make edits, but please don’t delete anything, comment out anything you intend to remove instead. Thanks!}
% \yt{TODO: 1. define the spin transport more carefully}
% \yt{TODO: 3. focus on introduction and selling our result, we can say: while there are some previous research demonstrating the quantum computing for magnetic susceptibility, we are the first one report the simulation of spin-current autocorrelation function, which is related to the spin conductivity.}
% \yt{TODO: 4. could include the heat current and energy current here ?}

\begin{abstract}
Understanding transport phenomena in quantum spin systems has long intrigued physicists due to their potential applications in spintronic devices and spin qubits.
Here, using a superconducting-qubit-based transmon device, we show that pre-fault-tolerant digital quantum simulation is reliable for studying transport phenomena via spin-current autocorrelation function (ACF).
While quantum simulations of the spin-spin ACF have been used to probe spin transport, methods based on the spin-current ACF have yet to be demonstrated due to their high gate cost, despite offering more direct information relevant to the transport properties.
Overcoming the resource constraints set by indirect measurement schemes like the Hadamard test, we showcase a direct measurement scheme that utilizes non-unitary operations, in particular mid-circuit measurements, to investigate spin transport for the 40-site 1D XXZ Heisenberg model in the near-ballistic, superdiffusive, and diffusive regimes. We successfully reproduce the expected power-law behavior in the superdiffusive regime and vanishing of the Drude weight in the diffusive regime.
% Evaluating transport properties of quantum spin systems is experimentally challenging not only due to fact that many are insulators, but also becasue of the intricate interplay between phonons and other carriers, and classically constrained due to exponential memory requirements with increasing system size. 
%Our implementation advances the study of spin currents and transport phenomena in quantum systems, contributing to the broader development of quantum materials and spintronic technologies.
\normalsize 
\end{abstract}

\maketitle

\textit{Introduction.}- Quantum spin systems are an ideal platform to investigate quantum entanglement \cite{latorre2003ground,verstraete2004entanglement,latorre2009short} and explore novel quantum phases. 
These systems host a variety of emergent quantum phenomena, such as quantum spin liquids \cite{kalmeyer1989theory, yunoki2006two, depenbrock2012nature}, topological magnons \cite{mcclarty2022topological}, and fractionalized excitations \cite{han2012fractionalized,dalla2015fractional}, and hold potential for applications in quantum sensing \cite{degen2017quantum}, spintronic devices \cite{wolf2001spintronics}, and spin-based qubits \cite{burkard2023semiconductor}, as energy and information can be transferred via the flow of spin angular momentum, known as the spin current.
A deeper understanding of how spin degrees of freedom behave and interact will advance both the fundamental physics and practical applications of quantum spin systems. 
Within quantum spin systems, understanding the transport behavior remains challenging as it involves the dynamics of systems.
% Thermal transport experiments serve as a critical tool for investigating quantum spin systems \cite{hess2003magnon, sun2003thermal, jin2003plane,  czajka2021oscillations}, particularly in insulating materials where charge transport is absent. 
% These measurements enable the study of thermal conductivity, which is used to characterize thermal and dynamical properties, under an applied temperature gradient. 
% However, characterizing spin transport in thermal experiments may be challenging due to the intertwined effects of magnon, phonons, and other carriers \cite{hlubek2012spinon, vandaele2017thermal}.

While prior works have investigated spin transport through spin-spin correlation functions \cite{keenan2023evidence, shi2024probing, kumaran2025quantum} and time-dependent magnetization \cite{rosenberg2024dynamics} on quantum computers, simulations of the spin-current autocorrelation function (ACF) remain largely unexplored, despite its significant relevance to spintronic devices and thermal transport experiments \cite{hess2003magnon, sun2003thermal, jin2003plane, czajka2021oscillations}.
This is primarily due to the higher gate cost associated with implementing the spin-current operator compared to the spin operator.
Although the spin-current ACF has been evaluated using classical methods \cite{herbrych2011finite,long2003finite,alvarez2002low,karrasch2013drude,steinigeweg2014spin,karrasch2012finite,karrasch2014real}, these approaches face limitations in handling many-body wave functions due to the memory issues.

Here, we propose a simple framework to evaluate spin transport based on spin-current ACF via digital quantum simulation.
We illustrate a direct measurement scheme that involves mid-circuit measurements (MCMs) based on Ref. \cite{mitarai2019methodology}, offering greater computational efficiency than the traditional Hadamard test. 
Compared to the Hadamard test, this direct measurement method does not require ancilla qubits and cross-qubit controlled gates. 
The number of circuits required to evaluate correlation functions for general $N$-particle systems scales as $\mathcal{O}(N)$, in contrast to $\mathcal{O}(N^2)$ for the Hadamard test.
We also demonstrate the utility of our protocol in evaluating different transport behaviors.
In addition, we further outline the general application of our proposed circuit for other correlation functions.
In summary, we demonstrate a spin-current evaluation method that is scalable for spin systems of sizes and geometries that are classically hard to deal with.

\begin{figure*}[htb]
\includegraphics[width=1\textwidth]
{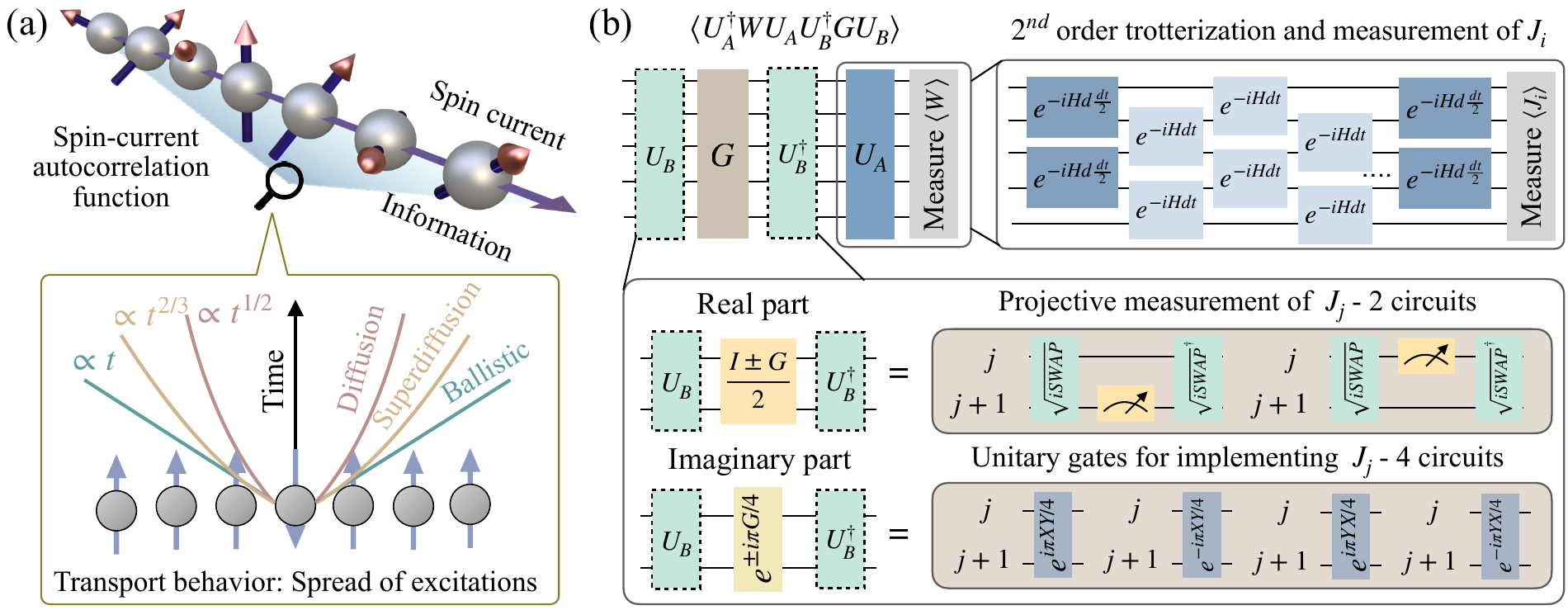}
\caption{\label{circuit_design}
(a) Illustration of spin current and the transport behavior of the spread of excitations, which is also illustrated in Ref. \cite{scheie2021detection}.
(b) Direct measurement protocol for evaluating the real and imaginary parts of $\langle U^{\dagger} W U G \rangle $ with only the gates without border (from Ref. \cite{mitarai2019methodology}) and $\langle U^{\dagger}_A W U_A U^{\dagger}_B G U_B \rangle $ with all the gates.
The unitary operator $U_B$ can be either used to transform the basis or encode the second time parameter.
The circuits in the box implement the measurement of the spin-current ACF.
The real-part evaluation is performed by implementing two different mid-circuit measurements while the imaginary-part evaluation is done by implementing four different unitary gates in circuits.
The Trotterized quantum circuit is detailed in Sec. \ref{implement_trotter}.
}
\end{figure*}

\textit{Spin-current ACF.}--
Consider the spin-1/2 XXZ Heisenberg model as it serves as the framework for studying magnetic interaction and spin dynamics. The Hamiltonian for the 1D chain as expressed by Eq. \ref{ham: chain}
\begin{equation}
\label{ham: chain}
 H = J \sum^L_{i=1} \left( S^X_i S^X_{i+1}+  S^Y_i S^Y_{i+1} + \Delta S^Z_i S^Z_{i+1} \right)
\end{equation}
Where $S^{X, Y, Z}_i$ are the spin-$\frac{1}{2}$ operators acting on $i^{th}$ spin, L is the length of spin chains, J is the coupling constant, and $\Delta$ is the anisotropy. For convenience, we set $J=$1 and consider various values of $\Delta$. The difference between the local spin currents $J_r$ and $J_{r-1}$ is defined via the continuity equation as the time derivative of the spin operator and combining this with Heisenberg's equation of motion, we have:
\begin{equation}
\label{continuity}
\frac{d S^Z_r}{dt} = i[H, S^Z_r] = - (J_{r} - J_{r-1}) 
\end{equation}
Using this, one can derive the local spin-current operator for a 1D chain as (See Sec. \ref{derive: spin current}):
\begin{equation}
\label{current: chain}
J_r = i[ S^Z_r, H_{r, r+1}] = J (S^X_r S^Y_{r+1} - S^Y_r S^X_{r+1})
% J_r = \sum_r i[ \sigma^Z_r, H_{r, r+1}] = J \sum_r (\sigma^X_r \sigma^Y_{r+1} - \sigma^Y_r \sigma^X_{r+1})
\end{equation}
Here, $r$ denotes the spin index, and $H_{r, r+1}$ represents the local Heisenberg Hamiltonian.

Instead of measuring the spin current, we aim to measure the spin-current ACF as it plays a central role in characterizing the transport behavior \cite{kubo1957statistical}.
Spin-current ACFs, which are two-time correlations between the spin current operators, are defined as
\begin{equation}
\label{cj(t)}
C_{J(t)J}= \sum_{i, j} \langle J_i(t) J_j \rangle /N
\end{equation} 
Different transport behaviors can be characterized by its frequency-dependent spectrum, decay profile, and Drude weight \cite{bertini2021finite}.
However, measuring the spin-current ACF is more computationally demanding than the spin current as it includes a time-dependent and a time-independent operator in the observable, which often requires the implementation of the Trotterized circuit acting on only one of the operators via the Hadamard test \cite{somma2002simulating}. 

\begin{figure*}[htp]
\includegraphics[width=1\textwidth]{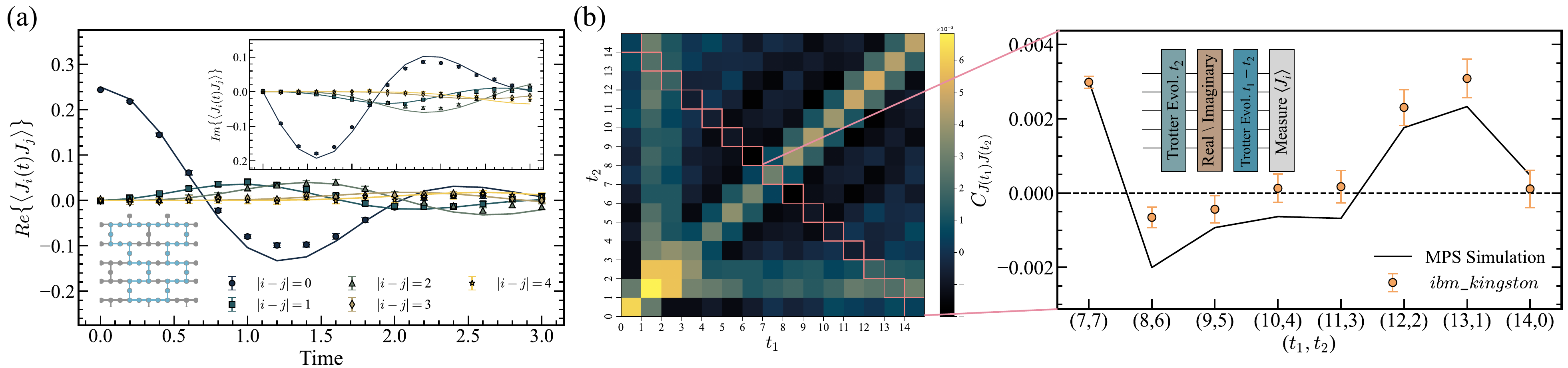}
\caption{\label{result_kingston}
(a) Quantum simulation of real and imaginary parts (inset) of local spin-current ACF $\langle J_i(t)J_j\rangle$ for a 40-qubit Neel state. 
The circuit layout on the $ibm\_kingston$ device is shown in the inset.
The solid lines are calculated via MPS simulation.
The Trotterization time interval is set to 0.2 with $\Delta = $2. 
(b) Demonstration of two-time spin-current ACF $C_{J(t_1)J(t_2)}$.
The heat map is generated using MPS simulation, and the quantum simulation is performed along the diagonal slice.
The quantum circuit for two-time spin-current ACF is shown in the inset.
The Trotterization time interval is set to 0.75 with $\Delta = $1.6. 
The error bar indicates the $95\%$ confidence intervals estimated by bootstrapping.
The results obtained from the $ibm\_kingston$ backend show good agreement with the MPS simulation for both one-time local and two-time spin-current ACF.
The deepest circuits for the real- and imaginary-part measurements have depths of 97 and 99, containing 1,882 and 1,880 two-qubit gates, respectively.
In the two-time spin-current experiment, the deepest circuit reaches a depth of 91 with 1,491 two-qubit gates.
}
\end{figure*}

\textit{Method}-- In principle, the Hadamard test \cite{somma2002simulating} can be used to evaluate the local spin-current ACF, $\langle U^\dagger J_i U J_j \rangle $ with $U=e^{-iHt}$, as it serves as a noise-resilient algorithm compared to qubitization \cite{low2019hamiltonian} and quantum singular value transformation \cite{gilyen2019quantum}.
However, in contemporary quantum devices, this indirect measurement method is less favorable because it requires the implementation of controlled-$J$ gates with ancilla qubits.
In the case of controlled-X gates, which are CNOT gates, both the number of CNOT gates and the circuit depth scale as $\mathcal{O}(N)$ for a CNOT gate applied across $N$ qubits to ancilla qubits \cite{baumer2024efficient}.
The cost of evaluating spin-current ACF via the Hadamard test quickly becomes unmanageable, as it involves several controlled-$J$ gates for different qubits.

To efficiently evaluate the $\langle U^\dagger J_i U J_j \rangle $ on quantum computers, we first develop a direct measurement method based on Ref. \cite{mitarai2019methodology}.
It has been shown that $\langle U^{\dagger} W U G \rangle $ can be measured without ancilla qubits, provided $U$ is a unitary matrix, 
$W$ and $G$ are Hermitian matrices, and $G^2 = I$.
The real part is evaluated using the $(I \pm G)/$2 gate while the imaginary part is obtained by employing the $e^{\pm i \pi G / 4}$ gate, as illustrated in Fig. \ref{circuit_design}(b) without the $U_B$ gate.
Notably, since $(I \pm G)/$2 is a projective and nonunitary operator, mid-circuit measurements (MCMs) are required for its implementation.
The expression $\langle U^{\dagger} W U G \rangle$ becomes a local spin-current ACF when $U$ is replaced by the time evolution operator $e^{-itH}$ and $W$ and $G$ are set to local current operator $J_i$,
\begin{equation}
    \langle U^{\dagger} W U G \rangle = \langle e^{i H t} J_i e^{-i H t} J_j \rangle = \langle J_i(t) J_j(0) \rangle
\end{equation}

Since all measurements must be performed on the Z basis, which is not the case for the current operator with MCMs, we extend the measurement protocol by introducing two unitary gates, $U_B$ and $U_B^{\dagger}$, as depicted in Fig. \ref{circuit_design}(b).
This circuit enables the implementation of the $\frac{I \pm J_r}{2}$ and $e^{\pm i \pi J_r /4}$ gates for measuring the real and imaginary parts, respectively. (See Sec. \ref{derivation}).
Besides, this circuit is also applicable for measuring two-time correlation functions as the time-evolution operator for the second time frame can be encoded in $U_B$ as well.
\begin{flalign}
\langle U_A^{\dagger} W U_A U_B^{\dagger} G U_B\rangle & = \langle e^{i H t_1} J_i e^{-i H t_1}e^{i H t_2} J_j e^{-i H t_2} \rangle \\
& = \langle J_i(t_1) J_j(t_2) \rangle 
\end{flalign}

The quantum circuit to measure the real part of the spin-current ACF is shown in Fig. \ref{circuit_design}(b).
To evaluate the expectation value $\langle J_i(t) J_j \rangle$, it's required to execute quantum circuits with two different MCMs.
The expectation values are computed based on the results from MCMs and final measurement following the strategy in Sec. \ref{expectation_real}.
To evaluate the imaginary part, four different circuits are required, and the expected value is determined based on the strategy in Sec. \ref{expectation_imag}.
Since the commuting local $\langle J_i \rangle$ can be measured simultaneously, it only requires two circuits ($\mathcal{O}(1)$) to measure all the local $J_i$ observable, compared to the N circuit ($\mathcal{O}(N)$) with Hardamard test.
Hence, to measure the spin-current ACF in Eq. \ref{cj(t)}, only $\mathcal{O}(N)$ circuits are required for direct measurement, compared to $\mathcal{O}(N^2)$ circuits in the Hadamard test.

For the quantum experiment, we perform our simulation on the IBM Heron processor $ibm\_kingston$, which comprises 156 fixed-frequency transmon qubits \cite{koch2007charge}, featuring heavy-hex connectivity.
We perform dynamical decoupling (DD) using the XY4 sequence \cite{maudsley1986modified} and employ 100 Pauli-twirled (PT) circuits \cite{wallman2016noise} with 128 shots to suppress the coherent noise.
We renormalize the noisy results using the depolarizing factor of the depolarizing channel, which is learned from a similar quantum circuit with a known outcome \cite{farrell2024scalable} (see Sec.~\ref{rescaling}).

\textit{Result.}--
We first validate our protocol using a 40-qubit Neel state under periodic boundary conditions.
The results for the real and imaginary parts of the local spin-current ACF $\langle J_i(t) J_j \rangle$ are presented in Fig.\ \ref{result_kingston}.
The agreement between the experimental results and the matrix-product-state (MPS) simulation demonstrates the effectiveness of our protocol and the noise modeling using a depolarizing channel.
Notably, the local spin-current ACF over long distances decays to zero for both the real and imaginary parts, indicating the properties of light-cone propagation.

In addition, we demonstrate the measurement of the two-time spin-current ACF $\langle J(t_1) J(t_2) \rangle$ using our protocol, as shown in Fig. \ref{result_kingston}(b), with the corresponding circuit displayed in the inset.
We evaluate the diagonal slice of the real part of the two-time spin-current ACF and validate our protocol by observing good agreement with MPS simulations using the initial state $|0\rangle^{\otimes 20} \otimes |1\rangle^{\otimes 20}$.
This initial state causes a spin flip which results in two domain walls being produced, and it will also be used for the transport experiment later. 
Notably, since the spin current is only generated at the domain boundary at $t=0$, we only need to evaluate $C_{J(t)J_d}$, where $J_d$ denotes the initial current across the domain boundary.

% We use the initial state $|0\rangle^{\otimes 20} \otimes |1\rangle^{\otimes 20}$, which will also be used for the transport experiment later, and measure the real part of the spin-current ACF.
% To validate our protocol, we evaluate the diagonal slice of the two-time ACF and obtain good agreement with MPS simulations.
% the two-time spin current ACF $\langle J(t_1) J(t_2) \rangle$ can also be measured using our protocol.
% Based on the Hermitian property of the spin-current operator, it only requires measurement in the case of $t_1 > t_2 $ to construct the entire heatmap in Fig. \ref{two_time}.
% We validate our protocol by measuring the diagonal slice and find good agreement with the MPS simulation.
% In practice, since the spin current is only generated at the boundary between $|0\rangle^{\otimes 20}$  and $|1\rangle^{\otimes 20}$ at $t=0$, which refers to domain wall, we only need to evaluate $C_{J(t)J_d}$, where $J_d$ denotes the initial current across the domain wall.

\begin{figure*}
\includegraphics[width=1\textwidth]{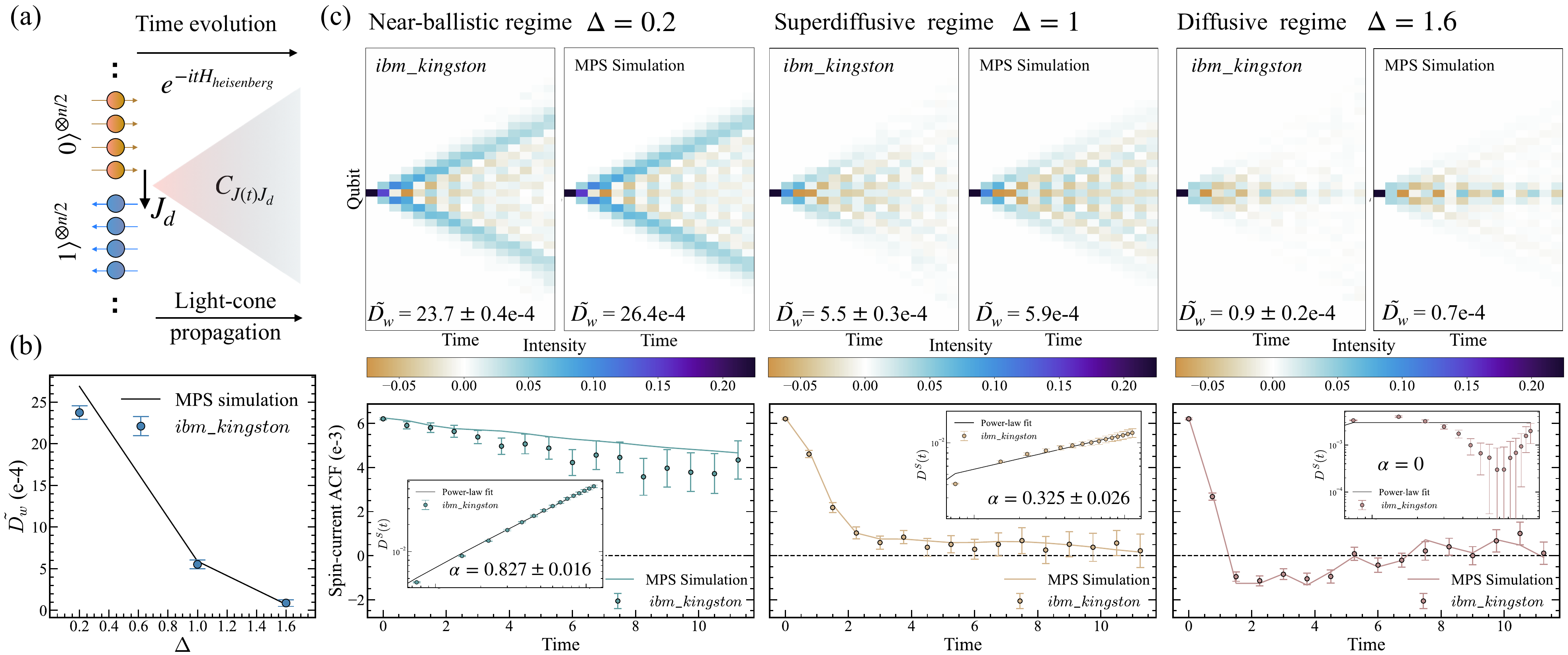}
\caption{\label{transport} 
Quantum simulation of a 40-qubit transport experiment.
(a) Illustration of the propagation of the spin-current ACF with an initial state of $|0\rangle^{\otimes n/2} \otimes |1\rangle^{\otimes n/2}$ .
% (b) Circuit layout on the $ibm\_kingston$ device.
(b) The Drude weight $\tilde{D}_w$ at $\Delta=$ 0,2, 1, and 1.6.
(c) Spin-current ACF at $\Delta=$ 0.2, 1, and 1.6.
The upper panel shows the local spin-current ACF $\langle J_i(t)J_d \rangle$ at each qubit site.
The bottom panel displays the spin-current ACF in Eq.~\ref{cj(t)}, with the corresponding diffusion coefficient $D^S(t)$ fitted with $t^{\alpha} $ shown with the log-log plot in the inset.
The error bar indicate the 95 $\%$ confidence intervals estimated by bootstrapping.
The Drude weight exhibits non-vanishing behavior in the ballistic regime and vanishing behavior in the diffusive regime. 
The fitted exponent $\alpha$ for the near-ballistic ($0.827$), superdiffusive ($0.325$), and diffusive ($0$) regimes falls within the theoretical range of $1$, $1/3$, and $0$, respectively. 
The deepest circuit has two-qubit gate depth of 99 and includes 1,789 two-qubit gates.
% \bp{To both this and fig2 add the two-qubit circuit depth and number of two-qubit gates.}
} 
\end{figure*}

Next, we investigate three distinct transport regimes by varying the anisotropy parameter $\Delta$ in the Heisenberg model. Ballistic transport arises for $0 < \Delta < $1, while diffusive behavior emerges when $\Delta > $1 \cite{gopalakrishnan2019kinetic}. At the isotropic point $\Delta = $1, the system exhibits superdiffusive transport \cite{ilievski2018superdiffusion}.
We aim to study the transport properties by evaluating the propagation behavior of the spin-current ACF, the Drude weight, and the scaling of the time-dependent diffusion coefficient. 
The spin-current ACF is expected to propagate within a linear light cone \cite{de2017spinon}, whose width is narrower in the diffusive regime than in the ballistic and superdiffusive ones \cite{ljubotina2017spin}.
The temperature-independent Drude weight $\tilde{D_w}$, defined as the long-time limit of the spin-current ACF integral,
\begin{equation}
\tilde{D}_w = \lim_{t \rightarrow \infty} \frac{1}{2t} \int_0^t C_{J(t')J} dt'
\end{equation}
 \cite{bertini2021finite}, 
% \bp{Should the integral be over t'?}
It is non-zero in the ballistic regime ($\Delta < $1) \cite{ilievski2017microscopic}, but vanishes in the diffusive regime ($\Delta > $1) \cite{bertini2021finite}.
At the superdiffusive point ($\Delta = $1), most theoretical predictions also indicate a vanishing Drude weight.
Finally, the time-dependent diffusion coefficient $D^S(t)$, expressed as the time integral of the spin-current ACF \cite{mcroberts2024parametrically} 
\begin{equation}
    D^S(t)= \int^t_0 C_{J(t')J} dt'
\end{equation}
follows a power-law scaling $t^\alpha$, where $\alpha = 1$ in the ballistic regime, $\alpha = 0$ in the diffusive regime, and $\alpha = 1/3$ \cite{gopalakrishnan2019kinetic, mcroberts2024parametrically} in the superdiffusive regime following the Kardar–Parisi–Zhang (KPZ) scaling \cite{kardar1986dynamic}.

We start with the initial state $|0\rangle^{\otimes 20} \otimes |1\rangle^{\otimes 20}$ and measure the real part of the spin-current ACF for systems evolved under $\Delta = 0.2$, $1$, and $1.6$, which theoretically correspond to the near-ballistic, superdiffusive, and diffusive regimes.
The illustration of the setup and the spread of the spin-current ACF are shown in Fig. \ref{transport} (a).
The quantum circuit is further optimized based on the linear light-cone structure, where only the qubits inside the light cone are measured, and unitary gates that do not contribute to the observable are removed (see Sec.~\ref{detail:transport}).

The propagation of the local spin-current ACF at each qubit site for these three transports is shown in the upper panel of Fig. \ref{transport}(c).
The width of the light-cone gradually decreases when transitioning from near-ballistic to diffusive regime, as the spin current becomes scattered and ceases to propagate effectively.
In the diffusive regime, the spin current decays quickly away from the domain wall with time evolution.
In contrast, in near-ballistic and super-diffusive regimes, the spin current can propagate to the qubits far from the domain wall, which is consistent with our expectation. 

The Drude weights are $(23.7 \pm 0.4)$ e-4 in the near-ballistic regime, and $(5.5 \pm 0.3)$ e-4 and $(0.9 \pm 0.2)$ e-4 in the superdiffusive and diffusive regimes, respectively, as shown in Fig. \ref{transport}(b).
The Drude weight in the near-ballistic regime remains consistent with a non-vanishing value on the order of $10^{-3}$, while the smaller value on the order of $10^{-5}$ in the diffusive regime reflects the vanishing characteristic of diffusive transport.
At the superdiffusive point, within our simulation time window and the finite length of the 1D chain, the spin-current ACF remains positive, indicating a non-zero Drude weight, as also observed in Ref. \cite{herbrych2011finite, karrasch2012finite, karrasch2013drude}.
Despite finite-size and finite-time effects, our results are consistent with those obtained from MPS simulations.

The spin-current ACF decays slowly in the near-ballistic regime, but more rapidly in the superdiffusive and diffusive regimes, as shown in the upper panel of Fig. \ref{transport}(c).
While the spin-current ACF remains greater than zero with a long tail similar to near-ballistic behavior in the superdiffusive regime, it quickly decays below zero and oscillates around zero in the diffusive regime, also observed in Ref. \cite{gobert2005real}.
By fitting the time-dependent diffusion coefficient, defined as the time integral of the spin-current ACF \cite{mcroberts2024parametrically}, to a power-law function $t^\alpha$, we obtained $\alpha = 0.827 \pm 0.016$, $0.325 \pm 0.026$, and $0$ for the near-ballistic, superdiffusive, and diffusive regimes, respectively, as shown in the inset of the bottom panel in Fig. \ref{transport}(c).
$\alpha = 0.827$ from the near-ballistic regime lies between the expected values of $1$ for ballistic transport and $1/3$ for superdiffusive transport, showing good agreement with theoretical predictions.
$\alpha = 0.325$ from the superdiffusive regime is consistent with the KPZ scaling exponent of $1/3$ at the isotropic point ($\Delta = 1$).
In the case of diffusive transport, no power-law scaling is observed, leading to the failure of fitting with $\alpha = 0$.
These quantitative agreements between the classical and quantum simulations of the spin-current ACF's light-cone, Drude weight, and power-law scaling of the diffusion coefficient all demonstrate that digital quantum simulations are reliable for studying spin transport.

\textit{Conclusion.}--
In this paper, we demonstrate a route to evaluate the transport behavior of many-body spin Hamiltonians by measuring the spin-current ACF. We first validate our circuit design by conducting experiments on quantum hardware with a 40-qubit N\'eel state, finding good agreement with MPS results for both the real and imaginary components.
In addition, we demonstrate the applicability of our method in evaluating the two-time spin-current ACF.
Next, we evolve a system composed of half spin-up and half spin-down under Heisenberg Hamiltonians with varying anisotropy and characterize distinct transport regimes through the light-cone property, Drude weight, and the power-law scaling of time-dependent diffusion coefficient. 

Our proposed measurement scheme can also be used to evaluate N-particle correlation functions, provided the operator can be expressed as a linear combination of weight-1 Pauli operators.
Two-time spin-spin and density-density correlations can be evaluated since they are weight-1 Pauli operators. 
Moreover, the retarded Green's function based on the local spin operator is also measurable.
For higher-weight Pauli operators, additional ancilla qubits are required to evaluate the real part \cite{mitarai2019methodology} while the imaginary part can be implemented without the restriction to weight-1 Pauli operators. 

A fundamental assumption of the utility of pre-fault- tolerant quantum computers is the promise of solving physically-motivated problems, and Hamiltonian simulation is one of the most promising of these physically-motivated problems. 
In this work, we establish the reliability of digital quantum simulation in modeling and studying non-equilibrium physics in spin systems.  
Our work advances the study of spin currents and transport phenomena and contributes to the broader development of quantum materials and spintronic devices. While the XXZ Heisenberg model considered here is classically tractable to simulate, we are hopeful that with advances in error suppression, mitigation and detection methods as well as quantum hardware improvements, more sophisticated transport problems will soon be within reach.  

\textit{Acknowledgments}--
The authors thank Keerthi Kumaran for constructive discussions and valuable feedback.
The work at the University of Illinois at Urbana-Champaign (UIUC) is supported by the Taiwan UIUC Scholarship under the official memo No.\ 1100063269M and by the IBM Illinois Discovery Accelerator Institute (IIDAI). The work performed at IBM and Purdue University is supported by the United States Department of Energy, the National Quantum Initiative Science Research Center, and the Quantum Science Center, managed by Oak Ridge National Laboratory. 
YTL also acknowledges the support of the IBM Summer Internship Program. 
This work made use of the Illinois Campus Cluster, a computing resource that is operated by the Illinois Campus Cluster Program (ICCP) in conjunction with the National Center for Supercomputing Applications (NCSA) and which is supported by funds from UIUC. 

\bibliographystyle{apsrev}
\bibliography{./main.bib}
% \end{document}

\beginsupplement
\newpage
%create title
\begin{center}

\textbf{\large Digital quantum simulation of spin transport \\\vspace{0.3 cm}}
% \textbf{Supplementary Information}

Yi-Ting Lee $^{1}$, Bibek Pokharel$^{2,3}$, Jeffrey Cohn$^{3,4}$ , Andr\'e Schleife$^{1,5}$, and   Arnab Banerjee$^{3,6}$

\small

$^1$\textit{Department of Materials Science and Engineering, University of Illinois at Urbana-Champaign, Urbana, IL 61801, USA}

$^2$\textit{IBM Quantum, IBM Thomas J. Watson Research Center, Yorktown Heights, NY, USA}

$^3$\textit{Quantum Science Center, Oak Ridge National Laboratory, Oak Ridge, TN 37831, USA}

$^4$\textit{IBM Quantum, IBM Research – Almaden, San Jose, CA 95120, USA}

$^5$\textit{Materials Research Laboratory, University of Illinois at Urbana-Champaign, Urbana, IL 61801, USA}

$^6$\textit{Department of Physics and Astronomy, Purdue University, West Lafayette, IN 47906, USA}

\end{center}

\section{Spin current operator in 1D-spin 1/2 Heisenberg model} \label{derive: spin current}
Based on the continuity equation in Eq.~\ref{continuity} and the Heisenberg equation of motion, the difference between the spin currents $J_r$ and $J_{r-1}$ can be written as
$i[H, S^Z_r] + J_r - J_{r-1} = 0$.
Since the Heisenberg Hamiltonian we consider acts only on nearest-neighbor pairs, we only need to account for the local terms involving the $(r-1)$-th, $r$-th, and $(r+1)$-th qubits. The commutator can then be rewritten as
\begin{flalign*}
% \label{first}
i[H_{r,r+1}, S^Z_r] + i[H_{r,r-1}, S^Z_r] + J_r - J_{r-1} = 0
\refstepcounter{equation}\tag{\theequation}
\end{flalign*}
where $J_r$ is defined as $i[S^Z_r, H_{r, r+1}]$, and $J_{r-1}$ corresponds to the backward current at site $r$, given by $i[H_{r, r-1}, S^Z_r] = -i[S^Z_r, H_{r, r-1}]$.
Since the $ S^Z_r $ operators commute with the local $ZZ$ terms in the Heisenberg Hamiltonian, we only need to consider the commutator with the $XX$ and $YY$ interactions and set $J=1$, as shown in the following derivation.

% \begin{align}
\begin{flalign*}
% \label{J_r}
J_r &= i[S^Z_r, H_{r, r+1}] = i \langle [S^Z_r, S^X_r S^X_{r+1}]\rangle 
+ i\langle [S^Z_r, S^Y_r S^Y_{r+1}]\rangle  \\ 
&= i\langle [S^Z_r,S^X_r ]S^X_{r+1}\rangle +
i\langle S^X_r[
S^Z_r,S^X_{r+1}]\rangle
+ i\langle [S^Z_r,S^Y_r ]S^Y_{r+1}\rangle +
i\langle S^Y_r[
S^Z_r,S^Y_{r+1}] \rangle\\ 
& = i\langle iS^Y_r S^X_{r+1} \rangle + i\langle -iS^X_r S^Y_{r+1} \rangle \\ 
& = \langle S^X_r S^Y_{r+1} - S^Y_r S^X_{r+1}\rangle
\refstepcounter{equation}\tag{\theequation}
\end{flalign*}
% \end{align}

% and $J_{r-1}$ is considered as the the backward current at r-th site in the above equation.

\section{Derivation of direct measurement protocol} \label{derivation}
\subsection{Real part}
% Based on Ref. \cite{mitarai2019methodology}, the real and imaginary part of the $\langle U^{\dagger} W UG \rangle$ can be  evaluated by direct measurement.
% For the real part, it is realized by subtracting projective measurement of $(I \pm G)/2$ as: 
In Ref. \cite{mitarai2019methodology}, the real part of $\langle U^{\dagger} W UG \rangle$ can be evaluated as $\langle P U^{\dagger} W U P \rangle$ - $\langle M U^{\dagger} W U M \rangle$, where $P$ and $M$ are defined as $(I + G)/2$ and $(I - G)/2$, respectively.
Here, $W$ and $G$ are assumed to be Hermitian matrices, and $G^2 = I$.
The derivation is shown below:
\begin{flalign*}
\label{origianl_real}
\Re{\langle U^{\dagger} W UG \rangle} & = \frac{1}{4} (\langle (I+G)U^{\dagger}WU (I+G) \rangle - \langle (I-G)U^{\dagger}WU (I-G) \rangle ) \\
& = \frac{1}{2}(\langle IU^{\dagger}WU G\rangle + \langle GU^{\dagger}WUI \rangle) \\
& = \frac{1}{2}(\langle U^{\dagger}WU G\rangle + \langle U^{\dagger}WU G\rangle^{\dagger}) 
\refstepcounter{equation}\tag{\theequation}
\end{flalign*}
To adapt this measurement protocol for an operator that does not satisfy $G^2 = I$, we apply a unitary transformation to $G$ in Eq.\ \ref{origianl_real} using the unitary matrix $U_B$. 
We then prove that this approach can be used to calculate $ \Re{\langle U^{\dagger}_A W U_A U_B^{\dagger} G U_B \rangle} $ as follows:
% $\langle U^{\dagger}_A W U_A U_B^{\dagger} G U_B \rangle$, one can simply sandwich the projector measurement $(I \pm G)/2$ by unitary transformation of $U_B$ as $U_B^{\dagger} (I \pm G) U_B/2 $, which can be achieved by sandwich the projector gate by twounitary gates. 
% The real part will be evaluated as the difference of $U_B^{\dagger} (I \pm G) U_B/2 $ result as follow: 
\begin{flalign*}
\label{transformation_real}
&\frac{1}{4} (\langle U_B^{\dagger}(I+G)U_B U_A^{\dagger}WU_A U_B^{\dagger}(I+G)U_B\rangle - 
\langle U_B^{\dagger}(I-G)U_B U_A^{\dagger}WU_A U_B^{\dagger}(I-G)U_B\rangle) \\
& \ \ \ \ \ \ \ \ \ \ =\frac{1}{2} (\langle U_B^{\dagger} I U_B U_A^{\dagger}WU_A U_B^{\dagger}G U_B\rangle 
+ (\langle U_B^{\dagger} G U_B U_A^{\dagger}WU_A U_B^{\dagger} I U_B\rangle \\
& \ \ \ \ \ \ \ \ \ \ = \frac{1}{2} (\langle U_A^{\dagger}WU_A U_B^{\dagger}G U_B\rangle + \langle U_B^{\dagger} G U_B U_A^{\dagger}WU_A\rangle) \\
& \ \ \ \ \ \ \ \ \ \ = \frac{1}{2} (\langle U_A^{\dagger}WU_A U_B^{\dagger}G U_B\rangle +\langle U_A^{\dagger}WU_A U_B^{\dagger}G U_B\rangle^{\dagger}) \\
& \ \ \ \ \ \ \ \ \ \ = \Re{\langle U^{\dagger}_A W U_A U_B^{\dagger} G U_B \rangle} 
\refstepcounter{equation}\tag{\theequation}
\end{flalign*}
$U_B$ can be used to unitarily transform the operator, ensuring it satisfies the $G^2 = I$ condition.
Moreover, $U_B$ can be combined with the time evolution operator, which is also a unitary matrix.
By replacing $U_A$ with the time evolution operator at $t_1$ and introducing the time evolution operator at $t_2$ within $U_B$, one can evaluate the two-time correlation function $\Re\langle W(t_1) G(t_2) \rangle$ and also satisfy $G^2 = I$ for circuit implementation.

\subsection{Imaginary part}
As for the imaginary part of $\langle U^{\dagger} W U G \rangle$, it can be evaluated using $\langle e_p U^{\dagger} W U e_m \rangle - \langle e_m U^{\dagger} W U e_p \rangle$, where $e_m$ and $e_p$ correspond to $e^{-i\pi G/4}$ and $e^{+i\pi G/4}$, respectively. This requires implementation with two separate circuits.
Here, $W$ and $G$ are assumed to be Hermitian matrices, and $G^2$ does not need to equal $I$ as long as $e^{\pm i \pi G/4}$ can be directly implemented on quantum circuits.
The derivation is shown as follows:
\begin{flalign*}
\label{original_imag}
\Im \{ \langle U^{\dagger}  W UG \rangle \} & = \frac{i}{2} (\langle e^{i \pi G/4} U^{\dagger}WU  e^{-i \pi G/4}\rangle - \langle e^{-i \pi G/4}U^{\dagger}WU  e^{i \pi G/4} \rangle ) \\
& = \frac{i}{2} (\langle (cos(\frac{\pi}{4})I + isin(\frac{\pi}{4})G) U^{\dagger} W U (cos(\frac{\pi}{4})I - isin(\frac{\pi}{4})G)\rangle - \\
&\ \ \ \ \ \ \ \ \ \ \ \langle (cos(\frac{\pi}{4})I - isin(\frac{\pi}{4})G) U^{\dagger} W U (cos(\frac{\pi}{4})I + isin(\frac{\pi}{4})G)\rangle ) \\ 
& = -i (\langle (cos(\frac{\pi}{4})I) U^{\dagger} W U (isin(\frac{\pi}{4})G)\rangle - \langle (isin(\frac{\pi}{4})G) U^{\dagger} W U (cos(\frac{\pi}{4})I)\rangle)  \\
& = \frac{1}{2} (\langle U^{\dagger} WU G \rangle- \langle GU^{\dagger}WU \rangle) \\ 
& = \frac{1}{2} (\langle U^{\dagger} WU G \rangle- \langle U^{\dagger} WU G \rangle^{\dagger})
\refstepcounter{equation}\tag{\theequation}
\end{flalign*}
To extend the measurement protocol, we also apply a unitary transformation to the $G$ component using the unitary matrix $U_B$, and demonstrate that this approach can be used to evaluate $\Im{\langle U^{\dagger}_A W U_A U_B^{\dagger} G U_B \rangle}$. 
The derivation is given as follows: 
\begin{flalign*}
\label{transformation_imag}
&\frac{i}{2} 
(\langle U_B^{\dagger} e^{i \pi G/4}  U_B U_A^{\dagger} W U_A  U_B^{\dagger} e^{-i \pi G/4} U_B \rangle
- \langle U_B^{\dagger} e^{-i \pi G/4}  U_B U_A^{\dagger} W U_A  U_B^{\dagger} e^{i \pi G/4} U_B \rangle) \\
& \ \ \ \ \ = \frac{i}{2} (\langle U_B^{\dagger}(cos(\frac{\pi}{4})I + isin(\frac{\pi}{4})G) U_B U_A^{\dagger} W U_A U_B^{\dagger} (cos(\frac{\pi}{4})I - isin(\frac{\pi}{4})G) U_B\rangle \\
&\ \ \ \ \ \ \ \ \ \ \ -\langle U_B^{\dagger}(cos(\frac{\pi}{4})I - isin(\frac{\pi}{4})G) U_B U_A^{\dagger} W U_A U_B^{\dagger} (cos(\frac{\pi}{4})I + isin(\frac{\pi}{4})G) U_B\rangle ) \\ 
& \ \ \ \ \ = -i (\langle U_B^{\dagger}(cos(\frac{\pi}{4})I)U_B U_A^{\dagger} W U_A U_B^{\dagger}(isin(\frac{\pi}{4})G) U_B \rangle \\
& \ \ \ \ \ \ \ \ \ \ \ - \langle U_B^{\dagger}(isin(\frac{\pi}{4})G)U_B U_A^{\dagger} W U_A U_B^{\dagger}(cos(\frac{\pi}{4})I)\rangle U_B)  \\
& \ \ \ \ \ = \frac{1}{2} (\langle U_A ^{\dagger} WU_A U_B^{\dagger} G U_B \rangle  - \langle U_B^{\dagger}G U_B U^{\dagger}_A WU_A  \rangle) \\ 
& \ \ \ \ \ = \frac{1}{2} (\langle U_A ^{\dagger} WU_A U_B^{\dagger} G U_B \rangle - \langle U_A ^{\dagger} WU_A U_B^{\dagger} G U_B \rangle^{\dagger})\\
& \ \ \ \ \ = \Im{\langle U^{\dagger}_A W U_A U_B^{\dagger} G U_B \rangle} 
\refstepcounter{equation}\tag{\theequation}
\end{flalign*}

\section{General implementation of measuring the  spin-current autocorrelation function} 

The following derivation is based on the Pauli operators $\sigma^{X, Y, Z, I}$, which can be directly implemented without considering the prefactor, rather than the spin-1/2 operators $S^{X, Y, Z, I}$. 
The measured spin-current ACF should be rescaled by 1/16 based on the spin number of each spin-1/2 operator.

\subsection{Real part}\label{expectation_real}
To evaluate the real part of $\langle U^{\dagger} W U G \rangle$, one must compute
$\langle P U^{\dagger} W U P \rangle - \langle M U^{\dagger} W U M \rangle$, where $P$ and $M$ are defined as $(I + G)/2$ and $(I - G)/2$, respectively, as shown in the previous section.
To measure the real part of the spin-current ACF $\langle U_{t}^{\dagger} J_i U_{t} J_j \rangle$, we need to implement the $(I+J_j)/2$ and $(I-J_j)/2$ gates on the quantum circuit.
Since the spin-current operator $J_j$ is not in the $Z$-basis and involves projective measurements, we first decompose it as follows to clarify the implementation:

\begin{equation} 
\label{ZI_IZ} J_j = \sigma^X_j \sigma^Y_{j+1} - \sigma^Y_j \sigma^X_{j+1} = T^{\dagger}(\sigma^I_j \sigma^Z_{j+1} - \sigma^Z_j \sigma^I_{j+1})T 
\end{equation}
where $T$ is a unitary matrix implemented using the $\sqrt{iSWAP}$ gate on the quantum circuit.
In Eq. \ref{ZI_IZ}, both $\sigma^I_j \sigma^Z_{j+1}$ and $\sigma^Z_j \sigma^I_{j+1}$ satisfy $G^2 = I$, and the gates $(I \pm \sigma^I_j \sigma^Z_{j+1})/2$ and $(I \pm \sigma^Z_j \sigma^I_{j+1})/2$ satisfy the condition of being projective operators $P$, such that $P^2 = P$.
Furthermore, these projective operators are all in the $Z$-basis, which enables mid-circuit measurement.
The real part of the spin-current ACF can then be evaluated separately as:
\begin{flalign*} 
\label{total}  \langle U_{t}^{\dagger} J_i U_{t} J_j\rangle = \ \langle U_{t}^{\dagger} J_i U_{t} T^{\dagger} \sigma^I_j \sigma^Z_{j+1} T \rangle -
\langle U_{t}^{\dagger} J_i U_{t} T^{\dagger} \sigma^Z_j \sigma^I_{j+1} T \rangle &\refstepcounter{equation}\tag{\theequation} \end{flalign*}
where $U_B$ and $U_B^{\dagger}$ in Fig.\ref{circuit_design} correspond to $T$ and $T^{\dagger}$ in Eq. \ref{total}, respectively.
We then evaluate these two terms via mid-circuit measurement following Eq. \ref{transformation_real}, but for brevity, we only show the implementation of the second term, as the first term follows the same procedure.
For the second term in Eq. \ref{total}:

\begin{flalign*} 
\label{mid-circuit} \Re \langle U_{t}^{\dagger} J_i U_{t} T^{\dagger} \sigma^Z_j \sigma^I_{j+1} T \rangle = \langle T^{\dagger} \frac{(I+\sigma^Z_j \sigma^I_{j+1})}{2} T U_{t}^{\dagger} J_i U_{t} T^{\dagger} \frac{(I+\sigma^Z_j \sigma^I_{j+1})}{2} T \rangle \\ 
-\langle T^{\dagger} \frac{(I-\sigma^Z_j \sigma^I_{j+1})}{2} T U_{t}^{\dagger} J_i U_{t} T^{\dagger} \frac{(I-\sigma^Z_j \sigma^I_{j+1})}{2} T \rangle 
&\refstepcounter{equation}\tag{\theequation} \end{flalign*}
where $\frac{(I+\sigma^Z_j \sigma^I_{j+1})}{2}$ and $\frac{(I-\sigma^Z_j \sigma^I_{j+1})}{2}$ are implemented by measuring the $j$-th qubit mid-circuit to project onto 0 and 1, respectively.
The $T$ and $U$ terms are implemented using quantum gates.
This expectation value is evaluated with the quantum circuit in Fig.\ref{circuit_design} (b) as:
\begin{flalign*} 
\label{first_term} \Re \langle U_{t}^{\dagger} J_i U_{t} T^{\dagger} \sigma^Z_j \sigma^I_{j+1} T \rangle = \langle J_i \rangle_{P_0, j} - \langle J_i \rangle_{P_1, j} 
&\refstepcounter{equation}\tag{\theequation} 
\end{flalign*}
where ${P_0, j}$ and ${P_1, j}$ refer to the outcomes 0 and 1 from measuring the $j$-th qubit during the mid-circuit measurement.
The real part of the spin-current ACF is evaluated through a quantum circuit as
\begin{flalign*} \label{real_expval} \Re \langle U_{t}^{\dagger} J_i U_{t} J_j \rangle = \langle J_i \rangle_{P_0, j+1} - \langle J_i \rangle_{P_1, j+1} - (\langle J_i \rangle_{P_0, j} - \langle J_i \rangle_{P_1, j})
&\refstepcounter{equation}\tag{\theequation}
\end{flalign*}
where two separate mid-circuit measurements are required to measure the $(j+1)$-th and $j$-th qubits, as shown in Fig. \ref{circuit_design} (b).
In practice, since the spin-current operator $J_i$ has been unitarily transformed by $\sqrt{iSWAP}$ gates before the final measurement, we measure $(\sigma^I_i \sigma^Z_{i+1} - \sigma^Z_j \sigma^I_{j+1})$ following Eq. \ref{ZI_IZ}.
The expectation value will be 
\begin{flalign*} 
\label{real_measure} \Re \langle U_{t}^{\dagger} J_i U_{t} J_j \rangle = \langle \sigma^I_i \sigma^Z_{i+1}- \sigma^Z_i \sigma^I_{i+1}  \rangle_{P_0-P_1, j+1} - \langle \sigma^I_i \sigma^Z_{i+1}- \sigma^Z_i \sigma^I_{i+1} \rangle_{P_0-P_1, j}
&\refstepcounter{equation}\tag{\theequation}
\end{flalign*}

\subsection{Imaginary part}\label{expectation_imag}
To evaluate the imaginary part of $\langle U^{\dagger} W U G \rangle$, one must compute
$\langle e_p U^{\dagger} W U e_m \rangle - \langle e_m U^{\dagger} W U e_p\rangle$, where $e_m$ and $e_p$ correspond to $e^{-i\pi G/4}$ and $e^{+i\pi G/4}$, respectively, as shown in the previous section.
To measure the imaginary part of $\langle J_i(t)J_j \rangle$, the implementation of $e^{-i\pi J_j/4}$ and $e^{+i\pi J_j/4}$ is required. 
To simplify the process, we first decompose is as follows 
\begin{flalign*} 
\label{xy-yx}  \langle U_{t}^{\dagger} J_i U_{t} J_j\rangle = \ \langle U_{t}^{\dagger} J_i U_{t} \sigma^X_j \sigma^Y_{j+1} \rangle -
\langle U_{t}^{\dagger} J_i U_{t}  \sigma^Y_j \sigma^X_{j+1}  \rangle &\refstepcounter{equation}\tag{\theequation} \end{flalign*}
We then evaluate the imaginary part using an exponential gate, following Eq. \ref{transformation_imag}.
For brevity, we derive only the first term, as the derivation of the second term follows the same logic.
For the first term in Eq. \ref{xy-yx}
\begin{flalign*} 
\label{exponent_1} \Im \langle U_{t}^{\dagger} J_i U_{t} \sigma^X_j \sigma^Y_{j+1}\rangle = \frac{1}{2}(
\langle e^{i \pi \sigma^X_j \sigma^Y_{j+1}/4} U_{t}^{\dagger} J_i U_{t} e^{-i \pi \sigma^X_j \sigma^Y_{j+1}/4}  \rangle \\ 
-\langle e^{-i \pi \sigma^X_j \sigma^Y_{j+1}/4} U_t^{\dagger} J_i U_{t} e^{i \pi \sigma^X_j \sigma^Y_{j+1}/4}  \rangle )
&\refstepcounter{equation}\tag{\theequation} \end{flalign*}
For this exponential operator, since the X and Y bases can be easily transformed into the Z basis through the $H$ and $R_x(\pi/2)$ gates, which act as $U_B$, we only present the final form here.
Moreover, one can derive Eq. \ref{exponent_1} from Eq. \ref{original_imag}, as there are no constraints requiring $e^{-i\pi G/4}$ and $e^{+i\pi G/4}$ to be implementable projective operators, unlike the real part.
This expectation value evaluated with quantum circuit in Fig. \ref{circuit_design}(b) as: 
\begin{flalign*} 
\label{exponent_2} \Im \langle U_{t}^{\dagger} J_i U_{t} \sigma^X_j \sigma^Y_{j+1}\rangle = \frac{1}{2}(
\langle J_i\rangle_{ e^{-i \pi \sigma^X_j \sigma^Y_{j+1}/4}  }-\langle J_i  \rangle_{ e^{i \pi \sigma^X_j \sigma^Y_{j+1}/4}})
&\refstepcounter{equation}\tag{\theequation} \end{flalign*}
where the subscript refers to the implementation of unitary gate in Fig. \ref{circuit_design}(b).
The imaginary part of the spin-current ACF is evaluated through a quantum circuit as
\begin{flalign*} 
\label{exponent_3}  \Im \langle U_{t}^{\dagger} J_i U_{t} J_j\rangle = \frac{1}{2} [\langle J_i\rangle_{ e^{-i \pi \sigma^X_j \sigma^Y_{j+1}/4}  }-\langle J_i  \rangle_{ e^{i \pi \sigma^X_j \sigma^Y_{j+1}/4}} - (\langle J_i\rangle_{ e^{-i \pi \sigma^Y_j \sigma^X_{j+1}/4} }-\langle J_i  \rangle_{ e^{i \pi \sigma^Y_j \sigma^X_{j+1}/4}})]
&\refstepcounter{equation}\tag{\theequation} 
\end{flalign*}
where four different exponential gates need to be placed on the $j$-th and $(j+1)$-th qubits, as shown in Fig. \ref{circuit_design} (a).
In practice, same as the real-part measurement, we measure $\langle \sigma^I_i\sigma^Z_{i+1}-\sigma^Z_i\sigma^I_{i+1} \rangle$ instead of $\langle J_i \rangle$.
The expectation value is described as follow

\begin{flalign*} 
\label{imag_measurement}  \Im \langle U_{t}^{\dagger} J_i U_{t} J_j\rangle = \frac{1}{2} [ \langle \sigma^I_i \sigma^Z_{i+1}- \sigma^Z_i \sigma^I_{i+1}\rangle_{e^{-i \pi \sigma^X_j \sigma^Y_{j+1}/4}  }-\langle \sigma^I_i \sigma^Z_{i+1}- \sigma^Z_i \sigma^I_{i+1}  \rangle_{ e^{i \pi \sigma^X_j \sigma^Y_{j+1}/4}} \\
- (\langle \sigma^I_i \sigma^Z_{i+1}- \sigma^Z_i \sigma^I_{i+1}\rangle_{ e^{-i \pi \sigma^Y_j \sigma^X_{j+1}/4} }-\langle \sigma^I_i \sigma^Z_{i+1}- \sigma^Z_i \sigma^I_{i+1}  \rangle_{ e^{i \pi \sigma^Y_j \sigma^X_{j+1}/4}})]
&\refstepcounter{equation}\tag{\theequation} 
\end{flalign*}

\section{time evolution circuit} \label{trotter implement}

To implement the time evolution operator on quantum circuits, we perform the second-order trotterization via the Suzuki-Trotter formulas \cite{hatano2005finding} as follows.
\begin{equation}
\label{trotter}
 e^{-iHt} \approx  \left[\prod_{j=1}^N e^{-i \frac{t}{2n}h_j} \prod_{j=N}^{1} e^{-i \frac{t}{2n}h_j}\right]^n 
\end{equation}
where  $n$ and $t$ refer to the number of Trotter steps and the total time in the simulation, respectively.
In this work, $h_j$ denotes the minimal set of non-commuting operators that compose the Hamiltonian, which is a two-local XXZ Hamiltonian and can be expressed in terms of Pauli operators as $J(S^X_j S^X_{j+1} + S^Y_j S^Y_{j+1} + \Delta S^Z_j S^Z_{j+1})$.

\label{implement_trotter}
\begin{figure}[h!]
\includegraphics[width=0.95\columnwidth]{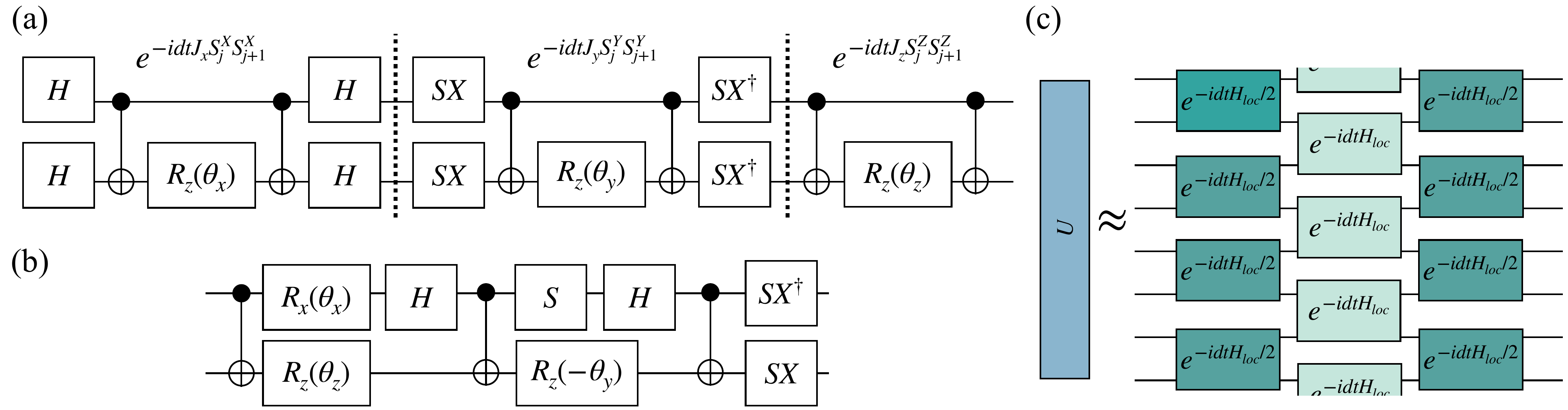}
\caption{\label{trotter_circuit} 
Circuit implementation for the local time evolution operator: (a) Direct implementation of time evolution for the $XX$, $YY$, and $ZZ$
interactions. (b) The optimized circuit for (a). 
The parameters $\theta_i =J_i*dt/2$, where $i \in x, y, z$.
(c) The circuit implementation of second-order trotterization, where $H_{loc}$ refers to local Hamiltonian $J(S^X_j S^X_{j+1} + S^Y_j S^Y_{j+1} + \Delta S^Z_j S^Z_{j+1})$.
} 
\end{figure}

The time-evolution gate of the Heisenberg model is implemented by the quantum circuit in Fig. \ref{trotter_circuit} (a).
As the $S^X_jS^X_{j+1}$, $S^Y_jS^Y_{j+1}$, and $S^Z_jS^Z_{j+1}$ operators commute with each other, $e^{-i t H_j}$ is implemented as a series of quantum gates.
One can further optimize the circuit in Fig. \ref{trotter_circuit} (a) to the one in Fig. \ref{trotter_circuit} (b), which has fewer CNOT gates.
When transpiling the quantum circuit with the highest level of optimization using \texttt{Qiskit} \cite{Qiskit}, the depths of the circuits for both the circuits in Fig.\ \ref{trotter_circuit} are identical.
The implementation of second-order trotterization is shown in Fig. \ref{trotter_circuit} (c).

% \yt[don't need this but need the collective measurement ]
\section{Noise Modeling: Depolarizing Renormalizing} 
\label{rescaling}

\begin{figure}[h!]
\includegraphics[width=0.5\columnwidth]{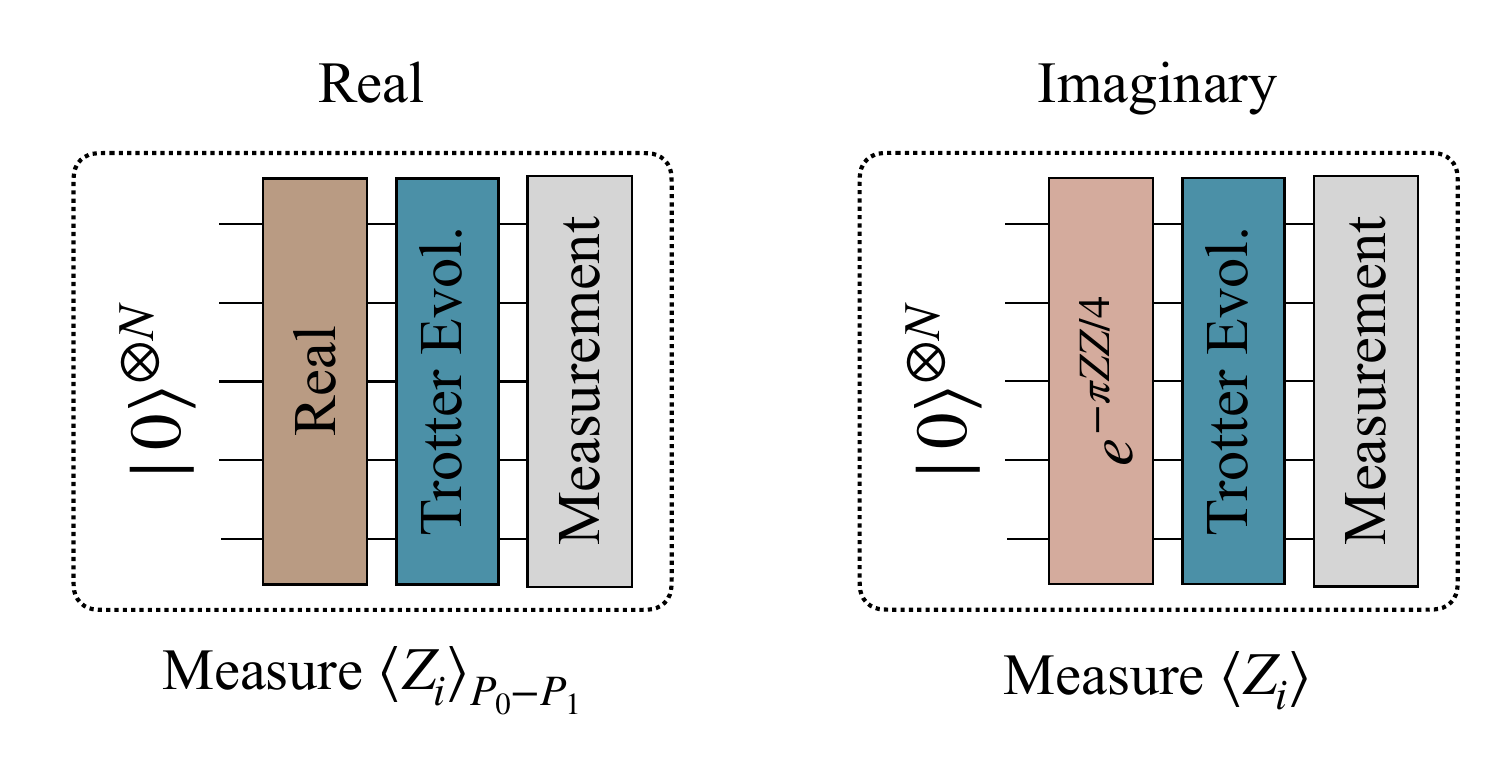}
\caption{\label{depo_circuit}
The quantum circuit used to learn the $(1-p)$ factor in Eq. \ref{eq:depo} for real and imaginary part. 
The real gate refers to the mid-circuit measurements for the real part in Fig. \ref{circuit_design}(b)
}
\end{figure}

To rescale our expectation values, we rely on operator decoherence renormalization \cite{farrell2024scalable}. In absence of classical verification of computation, rigorously proven error mitigation strategies like probabilistic error cancellation \cite{van2023probabilistic}, with clear error bounds on the reported expectation values would be preferred.
Operator renormalization is less resource intensive and works under the assumption that the noise channel can be approximated by the depolarization channel. 
In our quantum experiments, we perform dynamical decoupling (DD) using the XY4 sequence \cite{maudsley1986modified} and employ 100 Pauli-twirled (PT) circuits \cite{wallman2016noise} with 128 shots to mitigate the coherent noise.
The DD and PT circuit compilation is handled by \texttt{Qiskit} \cite{Qiskit}. 
With PT and DD, the noise channel can be approximated by
depolarization channel \cite{nielsen2010quantum}:  
\begin{equation}
\label{eq:depo_channel}
\varepsilon(\rho) = (1-p)\rho + p\frac{ I}{2^n}
\end{equation}
where $\rho$ is the density matrix and $p$ is the depolarization probability.
$n$ is the size of the system.
Since the expectation value of an operator $O$ is defined as $Tr[\rho O]$, one can derive the noisy observable $\langle \bar{O} \rangle$ (without any identity operator $I ^{\otimes n}$ in the $\bar{O}$) as  
\begin{equation}
\label{eq:depo}
\langle \bar{O} \rangle = (1-p) \langle O \rangle
\end{equation}
where $(1-p)$ is learned from similar circuits with a known answer.
For the real part, the circuit used to learn the depolarizing rate will be the same, except for the state preparation, For the imaginary part, we use the $e^{-i\pi ZZ/4}$ gate to approximate the noise pattern of the unitary gates in Fig. \ref{circuit_design}(b).
The circuits for learning the $(1-p)$ factor are shown in Fig. \ref{depo_circuit}.
With the $(1-p)$ factor, one can evaluate the noiseless result (See Sec. \ref{rescaling}).

% \begin{figure}[htp]
% \includegraphics[width=0.9\textwidth]{./SI/depolarization_real.png}
% \caption{\label{fidelity_real} 
% Strategy and results for estimating the $(1 - p)$ factor in Eq. \ref{eq:depo}, corresponding to the real part of the spin-current ACF. 
% The factor is determined by measuring $\langle Z_i \rangle_{P_0 - P_1}$, where $P_0$ and $P_1$ are the outcomes of mid-circuit measurements. 
% There are a total of four different factor maps, obtained from two distinct mid-circuit and reduced measurements.
% }
% \end{figure}
\subsection{Real part} \label{rescaling_real}
Based on Eq. \ref{eq:depo}, the depolarized results are obtained by rescaling the noiseless results by a factor of $1-p$, where p denotes the depolarizing rate.
For the spin-current ACF, evaluated using Eq. \ref{real_measure}, the depolarized results can be obtained from the following equation
\begin{flalign*} 
\label{depo_expval_real} 
\overline{\Re \langle U_{t}^{\dagger} J_i U_{t} J_j \rangle} = \langle \sigma^I_i \sigma^Z_{i+1} \rangle_{P_0-P_1, j+1} (1-p_{i+1})_{j+1} -\langle \sigma^Z_i \sigma^I_{i+1}  \rangle_{P_0-P_1, j+1}(1-p_{i})_{j+1} \\
- [\langle \sigma^I_i \sigma^Z_{i+1} \rangle_{P_0-P_1, j}(1-p_{i+1})_{j} - \langle \sigma^Z_i \sigma^I_{i+1} \rangle_{P_0-P_1, j}(1-p_{i})_{j}]
&\refstepcounter{equation}\tag{\theequation}
\end{flalign*}
This requires evaluating the depolarization rate of each qubit for two different circuits: one with a mid-circuit measurement on qubit $j$, and another on qubit $j+1$.
The  $(1-p)$ factor is evaluated by measuring $\langle Z_i \rangle_{P_0} - \langle Z_i \rangle_{P_1}$ with the initial state $|0\rangle^{\otimes N}$, where $P_0$ and $P_1$ correspond to the measurement results 0 and 1 from the mid-circuit measurement.
Since the observable is classically known to be 1, the measured result is the $(1-p)$ factor for  $i^{th}-$qubit.

\subsection{Imaginary part} \label{rescaling_imag}

% \begin{figure}[htp]
% \includegraphics[width=0.9\textwidth]{./SI/depolarization_imag.png}
% \caption{\label{fidelity_imag} 
% Strategy and results for measuring the $(1-p)$ factor in Eq. \ref{eq:depo} for the imaginary part of the spin-current ACF.
% The factor is determined by measuring $\langle Z_i \rangle$.
% There are two different factor maps, obtained from two distinct reduced measurements.}
% \end{figure}

The expectation value of the imaginary part is evaluated using Eq. \ref{imag_measurement}. For brevity in the following derivation, we denote $e^{i\pi \pm \sigma^X_j \sigma^Y_{j+1}/4}$ and $e^{i\pi \pm \sigma^Y_j \sigma^X_{j+1}/4}$ as $\pm XY$ and $\pm YX$, respectively.
The depolarized results can be evaluated as follows:
\begin{flalign*} 
\label{depo_expval_imag} 
\overline{\Im \langle U_{t}^{\dagger} J_i U_{t} J_j\rangle} = \frac{1}{2}[
\langle \sigma^Z_i \sigma^I_{i+1} \rangle_{-XY} (1-p_i)_{-XY}- \langle \sigma^I_i \sigma^Z_{i+1}\rangle_{-XY} (1-p_{i+1})_{-XY} \\
-(\langle \sigma^Z_i \sigma^I_{i+1} \rangle_{+XY} (1-p_i)_{+XY}- \langle \sigma^I_i \sigma^Z_{i+1}\rangle_{+XY} (1-p_{i+1})_{+XY})\\
-(\langle \sigma^Z_i \sigma^I_{i+1} \rangle_{-YX} (1-p_i)_{-YX}- \langle \sigma^I_i \sigma^Z_{i+1}\rangle_{-YX} (1-p_{i+1})_{-YX}) \\
+(\langle \sigma^Z_i \sigma^I_{i+1} \rangle_{+YX} (1-p_i)_{+YX}- \langle \sigma^I_i \sigma^Z_{i+1}\rangle_{+YX} (1-p_{i+1})_{+YX})]
&\refstepcounter{equation}\tag{\theequation} 
\end{flalign*}

In this research, we evaluate the factor $(1 - p)$ by implementing the $e^{-i \pi ZZ / 4}$ gate, which shares a structure similar to the four different exponential gates $\pm XY$ and $\pm YX$, shown in the imaginary gate in Fig. \ref{circuit_design}(b).
one can rewrite the Eq. \ref{depo_expval_imag} as:

\begin{flalign*}
\label{depo_expval_imag_rewrite} 
\overline{\Im \langle U_{t}^{\dagger} J_i U_{t} J_j\rangle} = \frac{1}{2}[
\langle \sigma^Z_i \sigma^I_{i+1} \rangle_{-XY} (1-p_i)_{-ZZ}- \langle \sigma^I_i \sigma^Z_{i+1}\rangle_{-XY} (1-p_{i+1})_{-ZZ} \\
-(\langle \sigma^Z_i \sigma^I_{i+1} \rangle_{+XY} (1-p_i)_{-ZZ}- \langle \sigma^I_i \sigma^Z_{i+1}\rangle_{+XY} (1-p_{i+1})_{-ZZ})\\
-(\langle \sigma^Z_i \sigma^I_{i+1} \rangle_{-YX} (1-p_i)_{-ZZ}- \langle \sigma^I_i \sigma^Z_{i+1}\rangle_{-YX} (1-p_{i+1})_{-ZZ}) \\
+(\langle \sigma^Z_i \sigma^I_{i+1} \rangle_{+YX} (1-p_i)_{-ZZ}- \langle \sigma^I_i \sigma^Z_{i+1}\rangle_{+YX} (1-p_{i+1})_{-ZZ})]
&\refstepcounter{equation}\tag{\theequation} 
\end{flalign*}
where ZZ denotes the  $e^{-i \pi ZZ / 4}$ gate.
We then determine the factor $(1 - p)_{-ZZ}$ by measuring $\langle Z_i \rangle$ with the initial state $|0\rangle ^{\otimes N}$, which is classically known to be 1.
% and the stategy is shown in Fig. \ref{fidelity_imag}, where the reduced measurement in Sec. \ref{reduced measurement} is also applied.
% Since both even-pair and odd-pair contributions are required to evaluate the total spin-current ACF, two distinct reduced measurements as described in Sec. \ref{reduced measurement}, need to be considered. The corresponding results are shown in Fig. \ref{fidelity_imag}.

% \begin{figure}[htp]
% \includegraphics[width=0.9\textwidth]{./SI/rescale_imag.png}
% \caption{\label{rescale_imag} 
% Imaginary part of the total spin-current ACF
% (a) Comparison between depolarized numerical and experimental results.
% (b) Comparison between rescaled experimental and numerical results.
% }
% \end{figure}

% Similar to the results for the real part, we depolarize the numerical data and rescale the experimental results.
% The comparison is shown in Fig. \ref{rescale_imag} (a) and (b), both of which again demonstrate good agreement.

\section{Implementation details for Neel state experiment} 
% \subsection{Reduced measurement}\label{reduced measurement}
\begin{figure}[htp]
\includegraphics[width=0.9\textwidth]{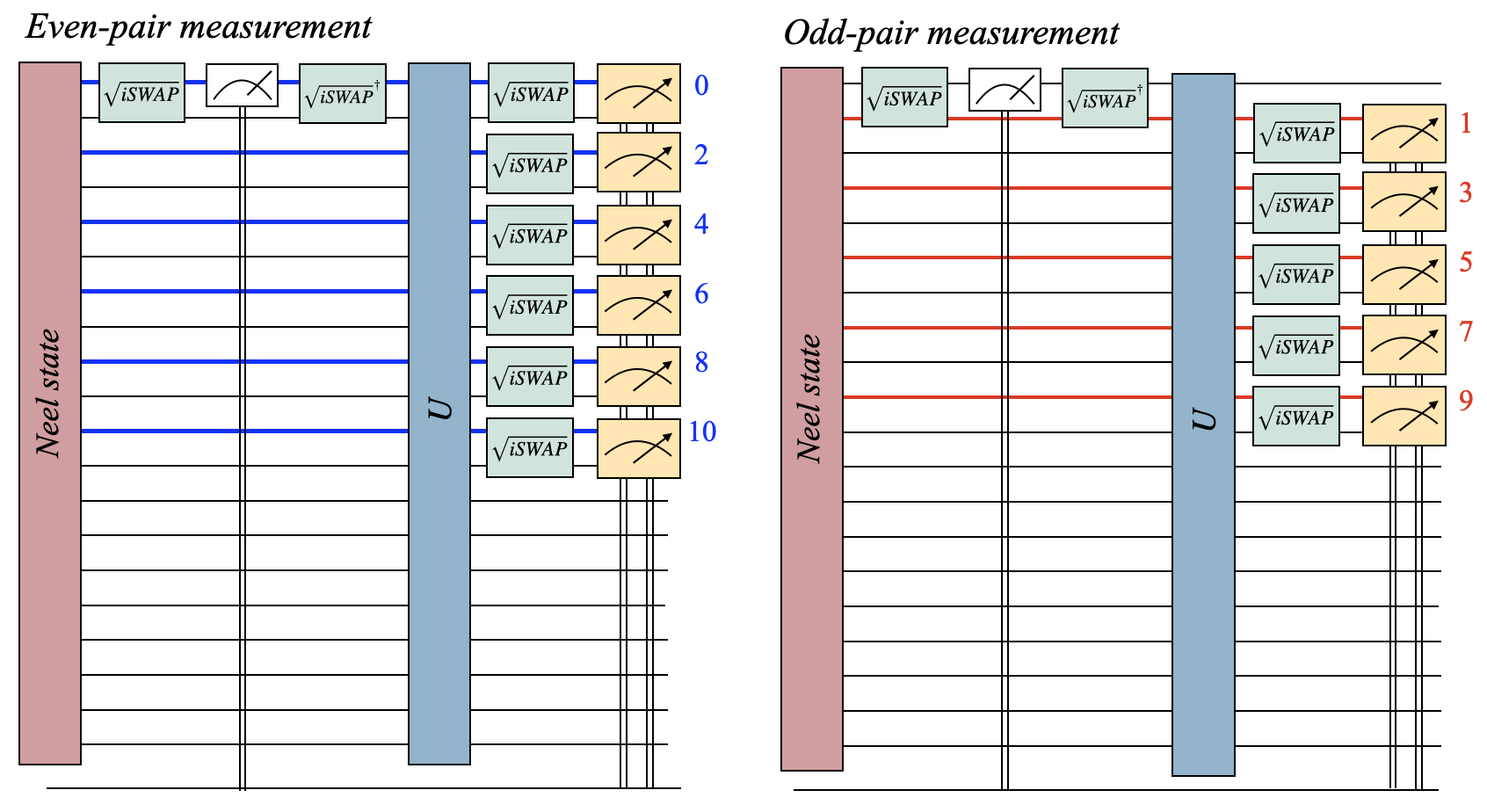}
\caption{\label{reduced_circuit} 
Example of reduced measurement protocol for evaluating all the real part spin-current ACF in Eq.\ \eqref{first_term} of the 1D 20-qubit N\'eel spin chain.
Based on translational symmetry and minimum-image convention, one only need to evaluate $\langle J_i(t) J_0 \rangle$, where $i = 0, 2, 4, \dots, n/2$ for even-pair measurements, and $i = 1, 3, 5, \dots, n/2 - 1$ for odd-pair measurements.
The measurement protocol for the imaginary part follows the same logic, where the mid-circuit measurement can be replaced with the imaginary gate in Fig. \ref{circuit_design}(b). 
% Since these circuits measure only half of the ACF, a total of 2 and 4 circuits are required to measure the real and imaginary parts, respectively. The bold circuit lines $i$ and $j$ indicate the target qubits for measuring the ACF. Time evolution gates are implemented as shown in Fig.\ \ref{trotterization} (b) and (c).
}
\end{figure}

For 1D spin chains, utilizing translational symmetry reduces the measurement problem to evaluating $\langle J_i(t) J_0 \rangle$ for $i = 0, 1, 2, \dots, n/2$, following the minimum-image convention. 
This is achieved by first placing the mid-circuit measurement and the exponential gate depicted in Fig.\ \ref{circuit_design} (b) on the first and second qubits, respectively.
Next, the measurement can be divided into even-pair measurements for $i = 0, 2, 4, \dots, n/2$ and odd-pair measurements for $i = 1, 3, 5, \dots, (n/2)-1$. 
Since all the local quantities can be evaluated using this measure strategy in Fig.\ \ref{reduced_circuit}, a total of four circuits are required to evaluate the real part, and eight circuits are needed to evaluate the imaginary part of all the spin-current ACFs, respectively.

\begin{figure*}[htp]
\includegraphics[width=0.9\textwidth]{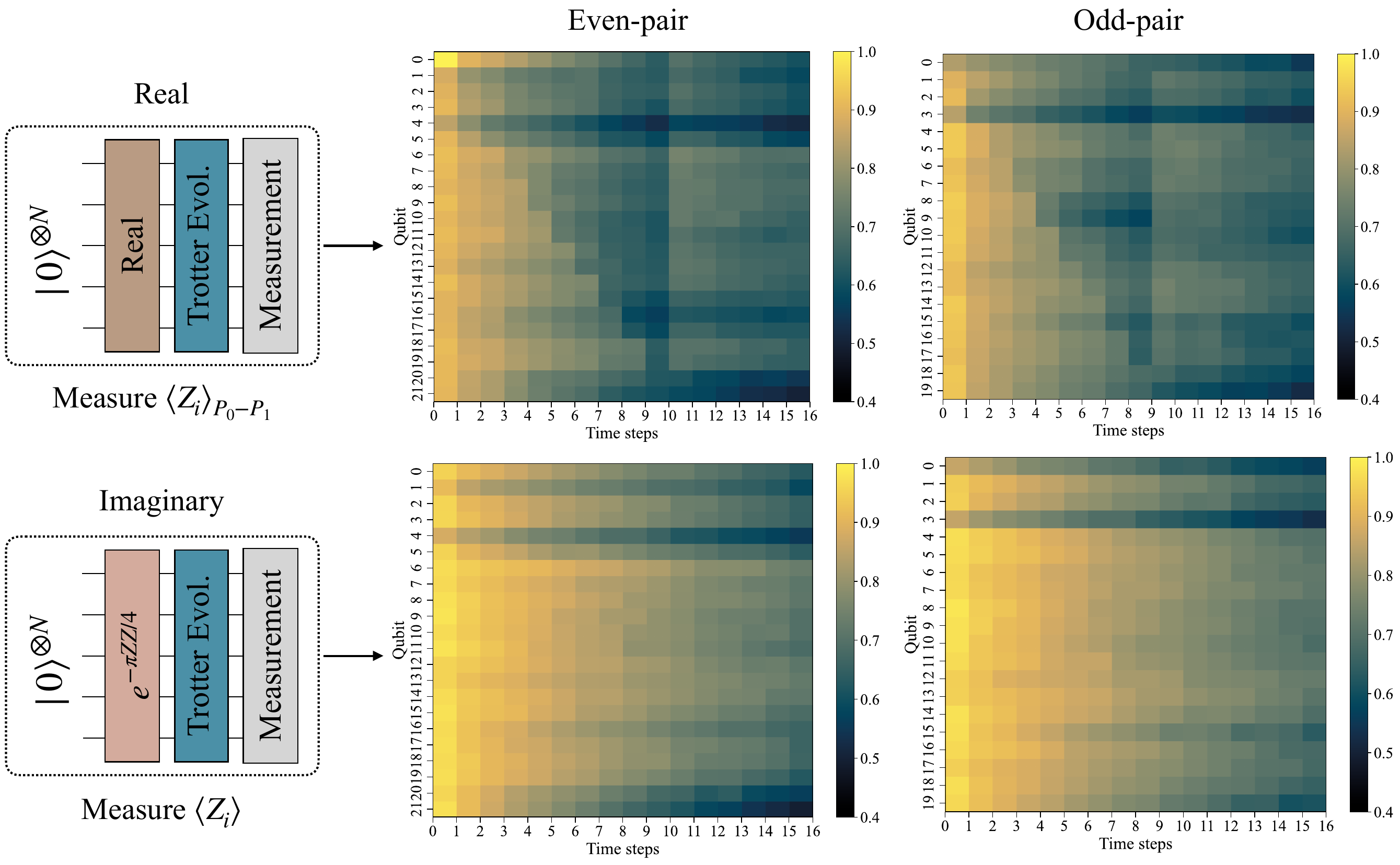}
\caption{\label{neel_depo}
The factor of $(1-p)$ in Eq. \ref{eq:depo} for evaluating the real and imaginary part spin-current ACF of a Neel state.
The circuit used to learn the $(1-p)$ factor are also shown in Fig. \ref{depo_circuit}.
}
\end{figure*}

Notably, for the Neel state, the absolute values of the observables for the two mid-circuit measurements (shown in Fig.~\ref{circuit_design}(b)) are the same.
Moreover, the absolute value of the observable for $e^{\mp i \pi XY/4}$ circuit is identical to that of $e^{\mp i \pi YX/4}$ circuit (shown in Fig.~\ref{circuit_design}(b)).
Therefore, we implement only one mid-circuit measurement and two different unitary circuits for the even- and odd-pair measurements, resulting in a total of two circuits for the real part and four circuits for the imaginary part.
The factor of $(1-p)$ in Eq. \ref{eq:depo} for the real part and imaginary part are shown in the Fig. \ref{neel_depo}.
% \subsection{Depolarizing rate}

\section{Implementation details for Transport experiment} 
\label{detail:transport}

\begin{figure*}[htp]
\includegraphics[width=0.9\textwidth]{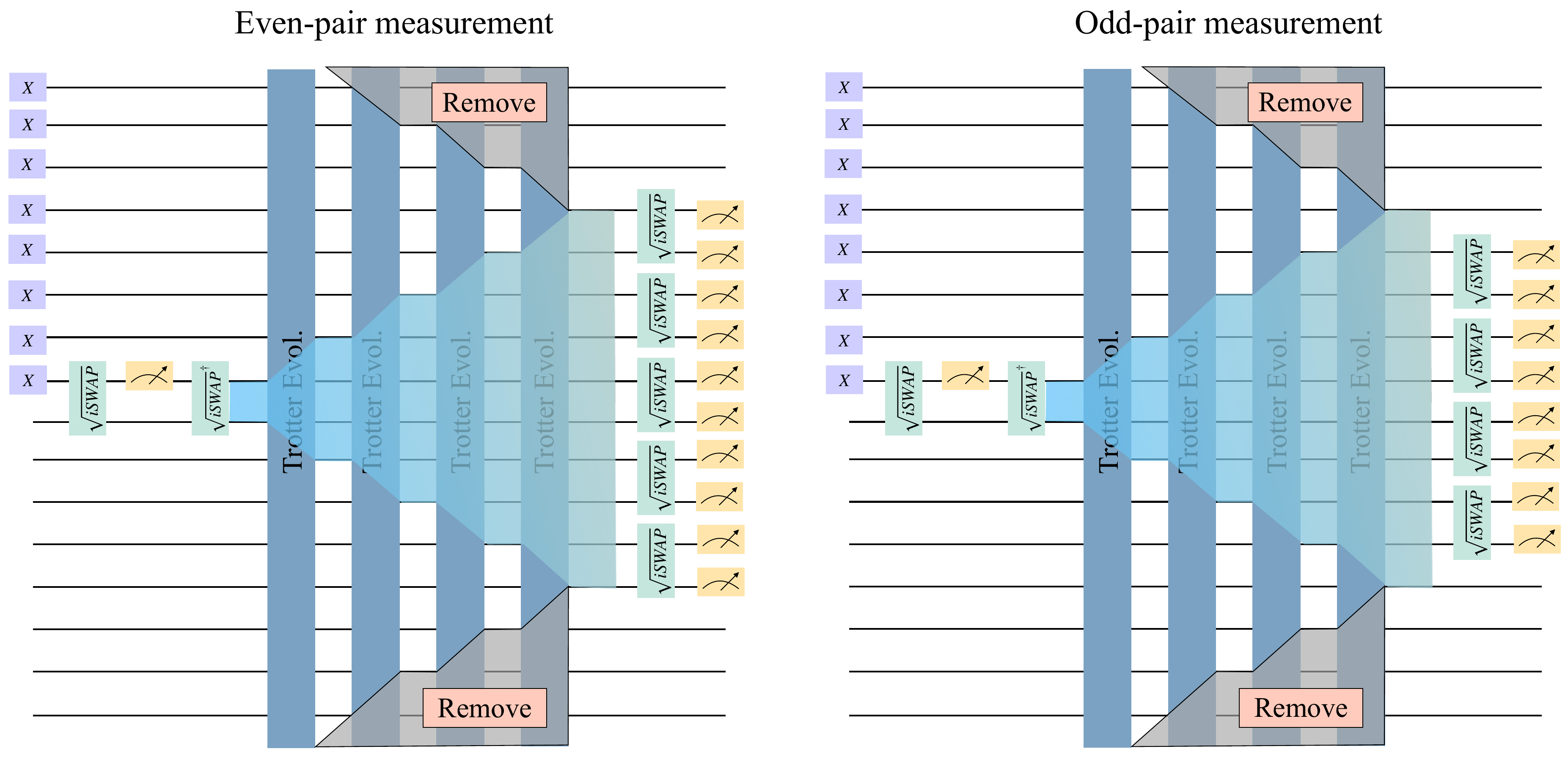}
\caption{\label{light_cone}
Illustration of circuit optimization using the light-cone property with even- and odd-pair measurements.
}
\end{figure*}

For the transport experiment, the spin-current ACF information is expected to propagate from the domain wall and remain confined within the linear light-cone region with the nearest-neighbor interaction.
To further optimize our quantum circuits, we leverage the fact that little to no information propagates outside the light cone. 
As a result, we restrict measurements to qubits within this region and reduce circuit complexity by eliminating gates located outside the effective light cone.
An illustration of this circuit optimization based on light-cone propagation is shown in Fig. \ref{light_cone}.
The velocity of the light cone is $1/dt = 1.33$ as the information propagates only to the nearest neighbor at each time frame, which is also straightforward for circuit compilation.
This velocity is greater than the maximum group velocity of spinons at $\Delta = 0.2\ (v_{max} = 1.124)$ \cite{de2017spinon}, and is close to that at $\Delta = 1\ (v_{max} = \pi/2)$ \cite{giamarchi2003quantum}.
% Magnons are typically the primary carriers of spin current in magnetically ordered systems, with a group velocity $J$, which is set to 1 in this study \cite{de2017spinon}.
% In addition, spinons can also carry spin current in one-dimensional systems. For $\Delta = 1$, the maximum spinon group velocity is approximately $\pi/2$, and it is below 1.2 for the other two cases considered \cite{de2017spinon}.
The light cone formed by the spin-current ACF is indeed narrower than our assumed linear light cone, further justifying the circuit optimization process.

\begin{figure*}[htp]
\includegraphics[width=0.9\textwidth]{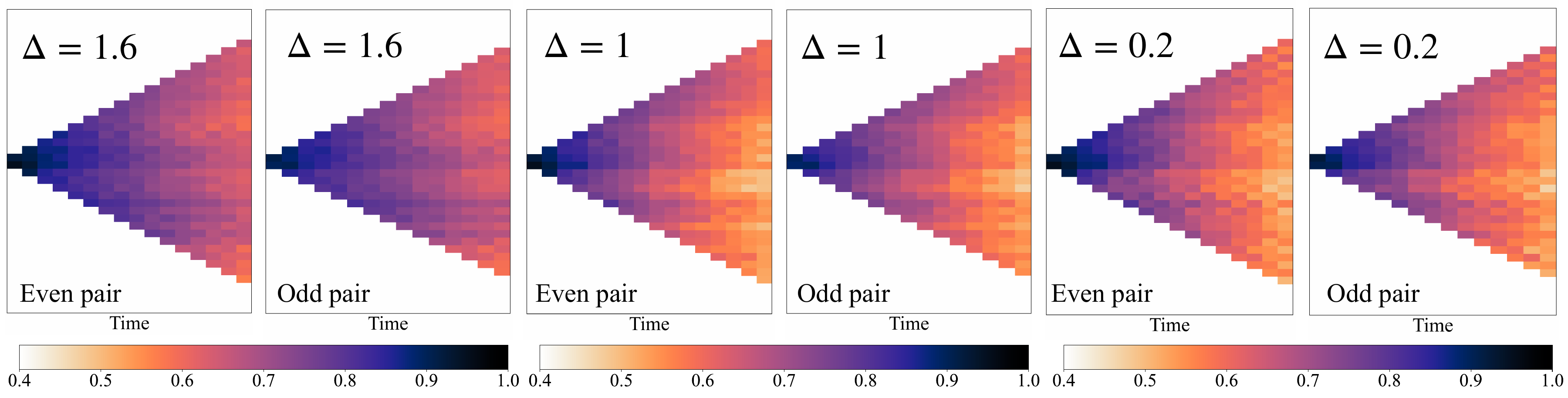}
\caption{\label{transport_1_p}
The factor of $(1-p)$ in Eq. \ref{eq:depo} for evaluating the real part spin-current ACF for three different cases in the transport experiment.
The circuit used to learn the $(1-p)$ factor are also shown in Fig. \ref{depo_circuit}.
}
\end{figure*}

To learn the $1 - p$ rescaling factor for the noisy results, we follow the same procedure as in the Néel state experiment. The $1 - p$ factor within the light-cone region for the three different transport cases is shown in Fig.~\ref{transport_1_p}.

% s expected to propagate within the linear light cone with the nearest neighbor interaction.

\section{Circuit and Hardware informaiton}
% For the 40-qubit N'eel state circuits, the number of two-qubit gates increases by 120 with periodic boundary condition.
% The deepest circuit for the real-part measurement has a depth of 97 and contains 1,882 two-qubit gates.
% For the imaginary-part measurement, the deepest circuit has a depth of 95 and includes 1,880 two-qubit gates.
% In the two-time spin-current experiment, the deepest circuit reaches a depth of 91 with 1,491 two-qubit gates.
% Lastly, for the transport experiment using the state $|0\rangle^{\otimes n/2} \otimes |1\rangle^{\otimes n/2}$, the deepest circuit has a depth of 99 and includes 1,789 two-qubit gates.
We perform the measurement using the $ibm\_kingston$ hardware, which consists of 156 fixed-frequency transmon qubits \cite{koch2007charge}, featuring heavy-hex connectivity.
The hardware information, including $T_1$, $T_2$, readout error mitigation, X-gate error, and CZ-gate error for performing the Neel state experiment, is summarized in the Table. \ref{table: neel cali}.
% \yt{waiting for credit being migrate, can't access the backend info}

% We perform the measurement using the $ibm\_kingston$ hardware, which consists of 156 fixed-frequency transmon qubits \cite{koch2007charge}, featuring heavy-hex connectivity and median $T_1$ and $T_2$ times of 216 $\mu s$ and 138 $\mu s$ , respectively. 
% The calibration data for each individual qubit in the 20-qubit layout, shown in Fig. \ref{result:real_marrakesh} (b), is summarized in Table. \ref{table: neel cali}.
% \section{Implementation for General correlation function}

\begin{table}[h]
\begin{tabular}{ccccccc} 
\toprule
Experiment &  & \shortstack{T1 (us)} & \shortstack{T2 (us)}  & 
\shortstack{Readout error}&
\shortstack{X error} &
\shortstack{CZ error (pair: error)} \\
% \colrule
\hline
\toprule
\centering
Neel state & Average &  217.42 &  134.52  &  0.017  &  0.00688  &  0.02957 \\ \hline
Neel state & Median  &  217.96 &  117.35  &  0.008  &  0.00027  &  0.002 \\ \hline
Two-time ACF & Average &  261.24 &  185.48  &  0.02  &  0.01331  &  0.03446 \\ \hline
Two-time ACF & Median &  268.0 &  155.86  &  0.009  &  0.00024  &  0.00197 \\ \hline
Transport & Average &  268.52 &  167.47  &  0.021  &  0.01317  &  0.03929 \\ \hline 
Transport & Median &  275.41 &  154.88  &  0.008  &  0.00025  &  0.00211 \\ \hline
\end{tabular}
\caption{\label{table: neel cali}
Calibration data for the 40-qubit Neel state experiment, two-time spin-current ACF experiment and the thermal experiment.}
\end{table}

\end{document}